\documentclass[useAMS,usenatbib]{mn2e}
\usepackage{epsfig}
\usepackage{amsmath}    % math package
\usepackage{amsfonts}
\usepackage{amssymb}
\usepackage{natbib}
\usepackage{myaasmacros}

\title[Measuring the BAO scale during reionisation]{Measurement of the baryonic acoustic oscillation scale in 21\,cm intensity fluctuations during the reionisation era} \author[K.J.  Rhook, P.M. Geil \& J.S.B. Wyithe]{Kirsty J. Rhook$^{1}$\thanks{krhook@ast.cam.ac.uk},
Paul M. Geil$^{2}$\thanks{pgeil@physics.unimelb.edu.au} 
\& J. Stuart B. Wyithe$^{2}$\thanks{swyithe@unimelb.edu.au} 
\\ $^{1}$Institute
of Astronomy, Madingley Road, Cambridge CB3 0HA\\ $^{2}$School of
Physics, University of Melbourne, Parkville, Victoria, Australia}

\begin{document}

\date{September 2nd, 2008}

\pagerange{\pageref{firstpage}--\pageref{lastpage}} \pubyear{2005}
  
\maketitle

\label{firstpage}

\begin{abstract}
  It has recently been suggested that the power spectrum of redshifted
  21\,cm fluctuations could be used to measure the scale of baryonic
  acoustic oscillations (BAOs) during the reionisation era. The
  resulting measurements are potentially as precise as those offered
  by the next generation of galaxy redshift surveys at lower
  redshift. However unlike galaxy redshift surveys, which in the
  linear regime are subject to a scale independent galaxy bias, the
  growth of ionised regions during reionisation is thought to
  introduce a strongly scale dependent relationship between the 21\,cm
  and mass power spectra. We use a semi-numerical model for
  reionisation to assess the impact of ionised regions on the
  precision and accuracy with which the BAO scale could be measured
  using redshifted 21\,cm observations. For a model in which
  reionisation is completed at $z\sim6$, we find that the constraints
  on the BAO scale are not systematically biased at $z \gtrsim
  6.5$. In this scenario, and assuming the sensitivity attainable with
  a low-frequency array comprising 10 times the collecting area of the
  Murchison Widefield Array, the BAO scale could be measured to within
  1.5~per~cent in the range $6.5 \la z \la 7.5$.
\end{abstract}


\section{Introduction}

The imprint of baryonic acoustic oscillations (BAOs) on the mass power
spectrum (PS) provides a cosmic yardstick that can be used to measure
the dependence of both the angular diameter distance and Hubble
parameter on redshift. The wavelength of a BAO is related to the size
of the sound horizon at recombination. Its value depends on the Hubble
constant, and on the dark matter and baryon densities. However, it
does not depend on the amount or nature of dark energy. Thus
measurements of the angular diameter distance and Hubble parameter can
in turn be used to constrain the possible evolution of dark energy
with cosmic time \citep[e.g.][]{eisenstein1998,eisenstein2002}.

Galaxy surveys have proved to be a powerful probe of the linear PS on
large scales and therefore provide a means to measure the BAO scale
\citep[e.g.][]{blake2003, seo2003, seo2005, glazebrook2005,
  hu2003,seo2007,angulo2008}. Indeed, analyses of the galaxy PS from
the SDSS and 2dFGRS surveys have uncovered a BAO signal
\citep[e.g.][]{cole2005,eisenstein2005,percival2007,okumura2008,gaztanaga2008,gaztanaga2008b},
providing incentive for deeper all-sky galaxy surveys, using
telescopes like the SKA\footnote{http://www.skatelescope.org},
Pan-STARRS\footnote{http://pan-starrs.ifa.hawaii.edu/public} and the
LSST\footnote{http://www.lsst.org}, to measure the BAO scale and hence
the dark energy equation of state more accurately.

Galaxy redshift surveys are best suited to studies of the dark energy
equation of state at relatively late times ($z \lesssim 3$) due to the
difficulty of obtaining accurate redshifts for a sufficiently large
number of high redshift galaxies. Although detection of the Integrated
Sachs-Wolfe effect puts some constraints on the integrated role of
dark energy above $z \sim 1.5$ [see, e.g., \citet{giannantonio2008}
and references therein], we currently have very limited information
about the nature of dark energy at high redshift. If dark energy
behaves like a cosmological constant, then its effect on the Hubble
expansion is only significant at $z\la1$ and becomes negligible at
$z\ga2$. In this case, studies of the BAO scale at low redshift would
provide the most powerful constraints.  However, as the origin of dark
energy is not understood we cannot presume a~priori which redshift
range should be studied in order to provide optimal constraints on
proposed models.  Probes of dark energy at higher redshifts have been
suggested.  These include measurements of the PS from a Ly$\alpha$
forest survey which could potentially be used probe the evolution of
dark energy through measurement of the BAO scale for redshifts as high
as $z\sim4$ \citep{mcdonald2007}. Similarly, studying the temporal
variation of high resolution quasar spectra may probe the evolution of
dark energy through the Sandage-Loeb test in the window $2<z<5$
\citep{corasaniti2007}.  Measurements of the BAO scale in the
reionisation era ($z>6$) are likely to be complimentary to the
constraints at lower redshifts provided by these techniques.

It has recently been suggested that observations of 21\,cm intensity
fluctuations might provide an additional avenue to measure the mass
PS \citep[e.g.][]{mcquinn2006, bowman2006,
mao_tegmark2008}, and that future low-frequency arrays could be used
to measure the BAO scale at a range of redshifts
\citep{wlg2008,chang2008,mao2008}. Redshifted 21\,cm surveys are
sensitive to neutral hydrogen regardless of whether it is part of a
resolved object. As a result, at high redshifts such surveys may be
more efficient in measuring the large-scale mass distribution than
galaxy redshift surveys, which must identify a very large number of
individual galaxies. Furthermore, spectroscopic emission line surveys
have the advantage of probing a precisely defined redshift interval,
which is unlikely to be true of a traditional galaxy survey
\cite[see][]{simpson2006}.

Planned low-frequency arrays such as the Murchison Widefield
Array\footnote{http://www.haystack.mit.edu/ast/arrays/mwa} (MWA) and
LOFAR\footnote{http://www.lofar.org} are designed to measure the PS of
21\,cm fluctuations at $z>6$ in order to probe the reionisation era.
\citet{wlg2008} have argued that while the first generation of
low-frequency arrays will not have sufficient sensitivity to precisely
determine the BAO scale, future extensions could have the sensitivity
to detect the BAO scale with promisingly small errors. \citet{wlg2008}
assumed a semi-analytic model which predicts a scale independent value
for the effective bias between the linear matter PS and the 21\,cm
PS. However the 21\,cm bias is expected to be strongly scale
dependent, especially toward the end of reionisation when many ionised
bubbles percolate through the intergalactic medium (IGM), creating an
excess of power on scales larger than the characteristic bubble size
\citep[e.g.][]{furlanetto2004}.  In this paper we explore whether the
scale dependence of the 21\,cm bias will compromise the ability of
redshifted 21\,cm observations to measure the BAO scale
\citep{mao2008}. We begin by reviewing the expected BAO signal and
21\,cm PS in Section~\ref{background}. We describe the observation of
the 21\,cm PS, including a discussion of the noise considerations in
Section~\ref{experiment}. In Section~\ref{fits} we describe our
fitting procedure for recovering the BAO scale and present our error
analysis, before concluding in Section~\ref{summary}. To calculate
these errors we adopt a set of cosmological parameters similar to
those derived from the third year WMAP data \citep{spergel2007}, namely
$\Omega_m = 0.24, \Omega_{\Lambda} = 0.76, h = 0.73, \sigma_8 =
0.0.76$ and $\Omega_b = 0.042$.

\section{Theoretical background}\label{background}

The sound speed at recombination $(s)$ determines the length scale of
the acoustic peaks in both the cosmic microwave background (CMB)
anisotropies and the linear matter PS. We refer to this length as the
BAO scale. The co-moving horizon size at decoupling $(r)$ may be
determined from knowledge of $\Omega_m$ and $\Omega_b$, and an
understanding of the physics governing a baryon-photon fluid
\citep{page2003}. \citet{komatsu2008} calculate this length scale to
be $146 \pm 1.8$\,Mpc for the cosmological parameters extracted from
the fifth year WMAP data. The corresponding angular scale $\theta_A =
r/D_A(z_{\rm rec})$ is closely related to the position of the first
acoustic peak in the CMB PS, which occurs at the scale of the mode
which has compressed once at the time of decoupling
\citep[see][]{page2003}. Following decoupling the same acoustic scale
is imprinted in the matter PS, albeit with a different phase. This
signature is expected to be preserved on large scales where non-linear
gravitational effects are small. Therefore the BAO scale may be used
as a standard ruler to test the geometry of the universe and the role
of dark energy.

\subsection{Theoretical expectations for the redshifted 21\,cm power spectrum}

Neutral hydrogen radiates at 21\,cm due to the hyperfine transition
between the singlet and triplet ground states. Assuming a contrast
between the kinetic temperature of the IGM $T_k$ and
the CMB temperature $T_{\rm CMB}$, and efficient coupling of the spin
temperature $T_s$ to $T_k$, the 21\,cm signal will mirror the underlying
density of neutral hydrogen.  The Ly$\alpha$ and X-ray flux emitted by
the first galaxies is expected to ensure these conditions hold during most
of the reionisation era \citep[see,
e.g.,][]{furlanetto2006}. A radio interferometer is sensitive to
fluctuations in the brightness temperature of neutral gas. For gas
of mean cosmic density the expected temperature brightness contrast 
in redshifted 21\,cm emission is
\begin{eqnarray}\label{deltaTb}\nonumber
  \delta T_b &=& \frac{T_b - T_{\rm CMB}}{(1+z)}(1-e^{-\tau})\\ \nonumber
  &\approx& 22~{\rm mK}~{x}_{\rm HI}\left(1-\frac{T_{\rm
        CMB}}{T_s}\right) 
  \left( \frac{\Omega_bh^2}{0.02}\right) \\\nonumber
  &&\hspace{15mm}\times\left[ \left( \frac{1+z}{7.5}\right) \left(
      \frac{0.24}{\Omega_m}\right)\right]^{1/2},\\
  & \equiv& \delta \bar{T}_b \bar{x}_{\rm HI}
\end{eqnarray}
where $\tau$ is the optical depth of the neutral gas to 21\,cm
radiation and $\bar{x}_{\rm HI}$ is the mass-weighted neutral fraction
of hydrogen in the IGM.

Allowing for fractional fluctuations in the baryonic matter density
$\delta(\Vec{x})$, ionised fraction $\delta_x(\Vec{x})$, and peculiar
gas velocity $\delta_v(\Vec{x}) =
\frac{1+z}{H(z)}\frac{dv_r}{dr}(\Vec{x})$ (where $\frac{dv_r}{dr}$ is
the peculiar velocity of the gas)\footnote{Variations in the local
  Ly-$\alpha$ flux may also result in fluctuations in the signal
  through $T_s$, \citep[e.g.][]{barkana2005a}}, the spatial dependence
of the brightness temperature fluctuations may be written
\citep[e.g.][]{mao2008}

\begin{eqnarray}\label{deltaTb_spatial}\nonumber
  \delta T_b (\Vec{x}) &=& \delta \bar{T}_b [1-(1-\bar{x}_{\rm HI})(1+\delta_x)](1+\delta)(1-\delta_v)\\
  &&\left(1-\frac{T_{\rm CMB}}{T_s}\right).
\end{eqnarray}

\noindent Assuming the peculiar velocity effect is described by the
linear theory result from \citet{kaiser1987}, \emph{i.e.}
$\delta_v(\Vec{k}) = - f \mu^2 \delta (\Vec{k})$ (note that $f \rightarrow
1$ in the high redshift limit) where $\mu = \Vec{k}.\Vec{n}$ denotes
the angle between the line of sight and the Fourier vector $\Vec{k}$, the
angle dependent 21\,cm power-spectrum may be written
\citep[e.g.][]{mao2008}

\begin{eqnarray}\label{21cmPS_angular}
  &&P_{21}(\Vec{k},z) = \delta \bar{T}_b^2[(\bar{x}_{\rm HI}^2 P_{\delta
    \delta} - 2 \bar{x}_{\rm HI}(1-\bar{x}_{\rm HI})P_{\delta x} +\\\nonumber
  &&(1-\bar{x}_{\rm HI})^2 P_{xx}) + 
  2 f \mu^2 (\bar{x}_{\rm HI}^2P_{\delta
    \delta} - \bar{x}_{\rm HI}(1-\bar{x}_{\rm HI})P_{\delta x})\\ 
  &&+ f^2 \mu^4\bar{x}_{\rm HI}^2P_{\delta \delta} ]. \nonumber
\end{eqnarray}

\noindent $P_{\delta \delta}$, $P_{x x}$ and $P_{\delta x}$ are the
linear matter power-spectrum, ionisation power-spectrum, and
ionisation-density cross power-spectrum respectively. In general the
21\,cm PS will display complex angle and redshift dependent
structure. The difference in the angular dependence modulating the
cosmological ($P_{\delta \delta}$) and astrophysical ($P_{\delta x},
P_{xx}$) components may be able to be exploited to constrain both
cosmological and astrophysical parameters
\citep[e.g.][]{barkana2005b,mao2008}.  The spherically averaged 21\,cm
PS may be more simply related to the underlying mass PS via
\begin{equation}
\label{P21Pm}
P_{21}(k,z) = b_{21}(k,z)^2 P_l(k,z) D^2(z),
\end{equation}
where $P_l$ is the primordial linear matter PS extrapolated to $z = 0$
calculated using CMBfast \citep{sz1996}, and $D(z)$ is the linear
growth factor between redshift $z$ and the present.  In
equation~(\ref{P21Pm}), the prefactor $b_{21}(k,z)$ is a term which is
analogous to galaxy bias (but which has units of mK). However in
contrast to galaxy redshift surveys, the value of $b_{21}$ can be
scale dependent, even in the linear regime, as a result of the
long-range, non-gravitational correlations introduced into the
ionisation structure of the IGM by the formation of HII
regions. Whilst peculiar velocity effects will be present,
\citet{seo2005} find that the BAO signal is preserved in the matter PS
at $z \sim 3$. Furthermore, \cite{seo2008} and \cite{shaw2008} show
that even the effect of non-linear redshift space distortions on the
spherically averaged matter PS may be able to be corrected to high
accuracy.

Equation~(\ref{P21Pm}) implicitly assumes that the gravitational
dynamics governing the large-scale structures probed are well
described by linearised fluid equations. Non-linear gravitational
dynamics will be increasingly important at low redshift. However at
the redshifts of interest here ($z>6$) non-linear evolution is
unlikely to play a significant role \citep[see, e.g.][]{crocce2008,
  mao2008}. \citet{mao_tegmark2008} find that the cut-off in the
maximum wave-number probed by redshifted 21\,cm PS in the range
$7<z<9$ has little affect on the constraints on cosmological
parameters.

\subsection{Model for the evolution of the 21\,cm signal}

In the early stages of reionisation when most of the hydrogen is still
neutral, the bias between the linear matter PS and the 21\,cm PS,
$b_{21}$, is well approximated by a constant with units of
temperature. However the bias is expected to depend on scale during
the process of reionisation since galaxies preferentially form in
overdense regions, resulting in early formation of ionised bubbles
from an overdense IGM. This effect can be modelled semi-analytically
\citep[e.g.][]{wyithe_morales2007}, or with full numerical
calculations of reionisation
\citep[e.g.][]{mcquinn2007,iliev2007,trac_cen2007}. Recently,
semi-numerical schemes have been devised which allow efficient
calculation of the large-scale ionisation field
\citep{mesinger2007,zahn2007, geil2008}. We use simulations computed
using the method described in \citet{geil2008} to model $b_{21}$.  We
refer the reader to the discussion of the semi-numerical modeling in
that paper, and to the discussion of the underlying semi-analytic
model for the reionisation of the IGM in
\citet{wyithe_morales2007}. Only a brief overview of the assumptions
and characteristics of the model(s) is included here.

\citet{wyithe_morales2007} describe a semi-analytic model for the
density and scale dependent ionisation fraction of the IGM. This model
assumes that a certain fraction of the mass in collapsed halos forms
galaxies, and that these galaxies produce ionising photons with some
efficiency.  The ionisation due to galaxies is opposed by the density
dependent recombination rate. An overdensity dependent rate of
ionisation may then be calculated as a function of scale. For the
examples shown in this paper, the model parameters are tuned such that
reionisation is completed by $z = 6$. \citet{wyithe_morales2007} use
this model to explore the 21\,cm brightness temperature contrast as a
function of scale and redshift.  \citet{geil2008} use these scale and
overdensity dependent predictions for the ionised gas fraction of the
IGM to generate realisations of the real-space density and ionisation
fraction of a region of the IGM. In this paper we use the
semi-numerical model to compute the 21\,cm bias $b_{21}$. We note that
our calculation does not include redshift space distortions
\citep*{barkana2005b} which are not captured by the model used in this
paper. The free parameters include the star formation efficiency,
escape fraction of ionising photons, and the minimum halo virial
temperature for star formation in neutral and ionised regions of IGM
(for which we take $10^4$\,K and $10^5$\,K respectively). The
semi-numerical simulation was performed in a cube with 256$^3$
resolution elements and of side length 3\,Gpc (co-moving).

\begin{figure*}
\vspace{5mm}
\includegraphics[width=16.5cm]{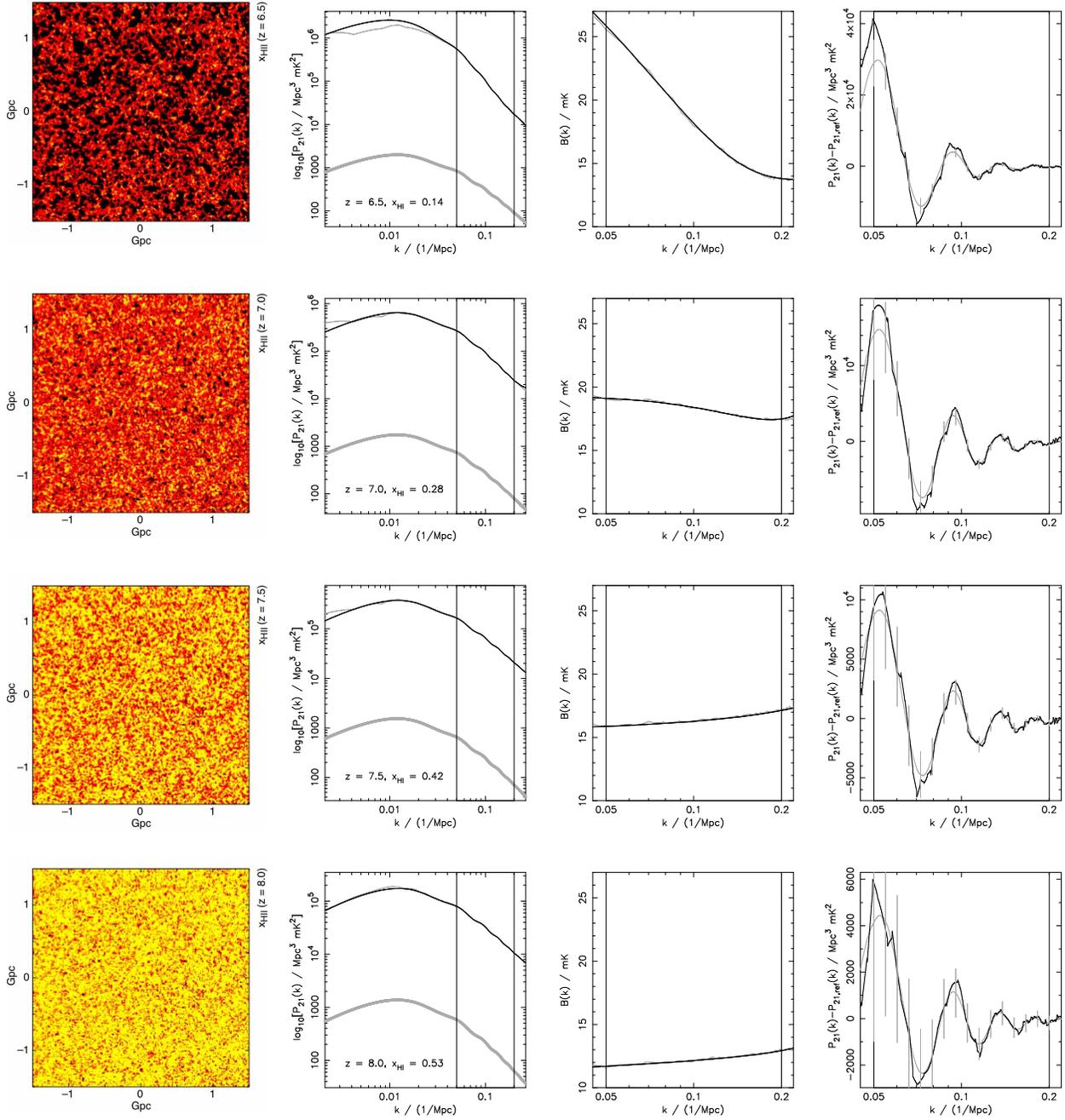} 
\caption{\label{fig1} Model power spectra and fits assuming a
third-order polynomial bias and no anomalous power. The left-hand
panels show the ionisation maps corresponding to redshifts of $z =
6.5$, 7.0, 7.5 and 8.0. The panels in the second column display the
corresponding model (\emph{grey}) and fitted (\emph{black}) 21\,cm PS,
and the linear matter PS (\emph{thick grey} line). The panels in the
third column show the fitted (\emph{black}) and model (\emph{grey})
scale dependent bias. In the far right-hand panels we plot the
difference between the model 21\,cm PS with and without baryons
(\emph{grey}), and the recovered 21\,cm PS with and without baryons
(\emph{black}), as described in the text. The \emph{grey} error bars
correspond to spherically averaged noise as described in
Section~\ref{experiment}. In each panel the vertical lines
indicate the range of $k$ values used for the fit.}
\end{figure*}

In Figure~\ref{fig1} we show the model 21\,cm PS at $z = 6.5$, 7, 7.5
and 8, corresponding to average volume weighted neutral fractions
$x_{\rm HI} = 0.14, 0.28, 0.42,0.53$.  The left-hand panels show the
ionised gas distribution from the semi-numerical simulations for a box
with side length 3\,Gpc (co-moving). The panels in the second column
show the spherically averaged model 21\,cm PS (grey lines). The panels
in the third column show the model 21\,cm bias (grey lines). The far
right-hand panels show the BAO component of the PS, defined as the
difference between the 21\,cm PS, and a reference 21\,cm PS computed
from a mass PS without baryons, $P_{l,nb}(k)$.  The model BAO
component, $b_{21}^2(k) D^2(z) [P_l(k)-P_{l,nb}(k)]$, is shown in
grey.  The model PS is reproduced in Figure 2.

\subsection{Fitting formula for 21\,cm power spectra}
 
Following \citet{seo2005} we parametrize the fit to the semi-numerical model of the spherically averaged 21\,cm PS as
\begin{equation}\label{model1}
P_{21,{\rm fit}}(k,z) = B^2(k)~D^2(z)~P_l(k/\alpha) + AP(k),
\end{equation}
where $AP(k) = c_0 + c_1k + c_2k^2$ is a term to describe anomalous
power and $B(k) = b_0 + b_1 + b_2k^2 + b_3k^3$ is the fitted bias in
units of mK.  The ``dilation'' or ``phase'' parameter $\alpha$ allows
for a dilation/contraction of the matter PS used to fit the 21\,cm
PS. In our case the input value of $\alpha$ is unity since we are
assuming we know the cosmological parameters governing
$P_l$. \citet{seo2005} note that for the parametrization the PS in
equation~(\ref{model1}), the errors in the known values of the
wave-number dilation parameter $\alpha$ directly translate into the
errors in the measured BAO scales. In Section~\ref{error} we calculate
the expected redshift dependent errors $\Delta \alpha$ for an
observational campaign which is described in Section~\ref{experiment}
below.

The anomalous power term allows for the distortion of the matter
power-spectrum due to non-linear mode coupling. \citet{seo2005} find
that by subtracting a smooth function ($AP$) from the matter PS
derived from their N-body simulations, they are able to recover the
shape of the matter PS and thus the BAO signal. Our semi-numerical
model is intrinsically linear, and so does not include anomalous
power. However we explore the possible effect of anomalous power on
the measurement of the BAO scale by including the correction term $AP$
in the fitted PS. Inclusion of this term approximates the effect of
degeneracy between anomalous power and scale dependent bias. In this
paper we present results using fits both with and without the $AP$
term.  Note that the degeneracy between the parameters governing the
anomalous power and bias terms means that the polynomial fits to these
functions cannot be interpreted physically when the $AP$ term is
included (see Section~\ref{fits}). However we find that this
degeneracy does not significantly hinder the determination of $\alpha$.

More detailed cosmological constraints may be obtainable if the
line-of-sight and transverse BAO scales could be measured
independently [see \citet{wlg2008} and references therein, and also
\citet{okumura2008}, \cite{gaztanaga2008} and
\cite{gaztanaga2008b} for recent constraints from galaxy
surveys]. Indeed, a low-frequency array is naturally suited to
measuring both the line-of-sight and transverse components of the
21\,cm PS (see Section~\ref{experiment}). Such an analysis would
require modelling of the peculiar velocity effects which give rise to
the angular dependence of the 21\,cm PS in
equation~\eqref{21cmPS_angular}. As mentioned, the model adopted for
the 21\,cm PS does not include peculiar velocities. However we note
that above fitting formula could be readily generalised to
characterise the ionisation and cross density-ionisation power-spectra
needed to describe the angular dependence in
equation~\eqref{21cmPS_angular}. This would allow constraints on the
dilation parameter, and therefore the BAO scale, both perpendicular
and parallel to the line-of-sight to be extracted from the 21\,cm
power-spectrum. The signal-to-noise ratio estimates in \citet{wlg2008}
suggest that the MWA5000 will be similarly sensitive to each of these
scales.

\section{Measurement of 21\,cm power spectra}
\label{experiment}

\begin{figure*}
\vspace{5mm}
\includegraphics[width=16.5cm]{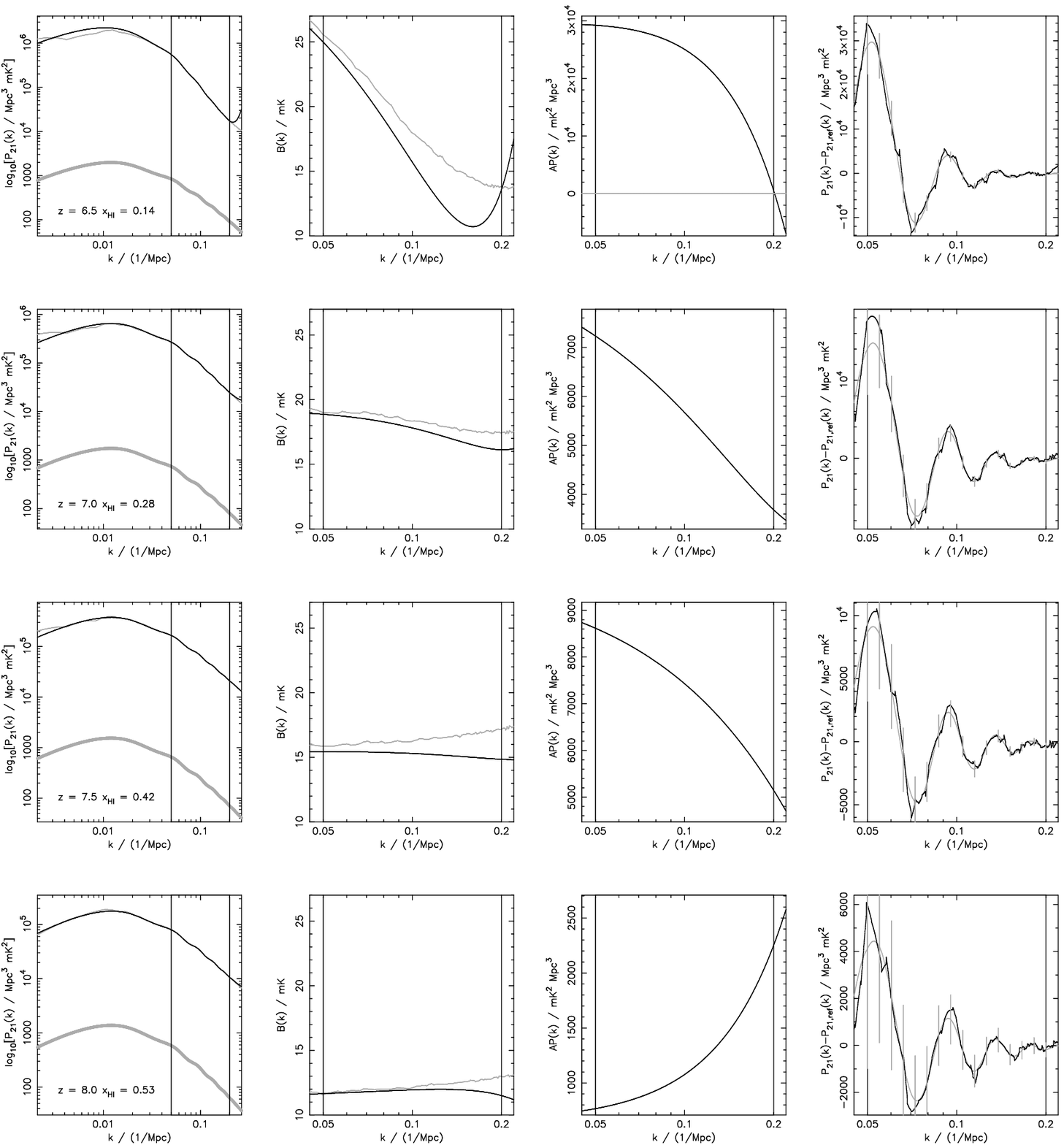} 
\caption{\label{fig2}Model power spectra and fits assuming a
third-order polynomial bias and quadratic anomalous power for the
same simulation data as Figure~1. The first column of panels display
the model (\emph{grey}) and fitted (\emph{black}) 21\,cm PS, and the
linear matter PS (\emph{thick grey} line) at redshifts of $z = 6.5$,
7.0, 7.5 and 8.0. The panels in the second column show the fitted
(\emph{black}) and model (\emph{grey}) scale dependent bias. The
anomalous power component of the fit is plotted in the third
column. In the far right-hand panels we plot the difference between
the model PS with and without baryons (\emph{grey}), and the recovered
PS with and without baryons (\emph{black}), as described in the text.
The \emph{grey} error bars correspond to spherically averaged noise as
described in Section~\ref{experiment}. In each panel the vertical 
lines indicate the range of $k$ values used for the fit.}
\end{figure*}

In this section we discuss the sensitivity of a low-frequency array to
the 21\,cm PS. The sensitivity to the PS depends on the array design
(e.g. total collecting area and antenna distribution) as well as the
observation strategy (e.g. total integration time).  We consider a
low-frequency array which would be a 10-fold extension of the MWA.  We
refer to this array as the MWA5000 \citep{mcquinn2006}. The MWA5000
would consist of 5000 antennae (with each antenna composed of a tile
of $4 \times 4$ cross-dipoles) with effective collecting area $A_e\sim
16 \lambda^2/4$\,m$^2$, where $\lambda = 0.21(1+z)$\,m is the observed
wavelength. The dipoles will be sensitive to a wide range of
frequencies which translate into redshifts spanning the expected
extent of reionisation. However, we assume that the receiver will
limit the bandpass to a selected $B_{\rm tot}=32$\,MHz window per
observation. We assume that the antennae are distributed with constant
density in a core of radius $80$\,m (within which the $u$-$v$ coverage
is very high), with the remainder distributed with a radially
symmetric density profile $\rho \propto r^{-2}$ out to a radius of
$1$~km (so that the maximum baseline is $\sim 2$\,km).

We use the prescription from \citet{bowman2006} for the 3d sensitivity
of an array to the 21\,cm PS. The array sensitivity is expressed in
terms of the wave-vector components which are orthogonal $k_{\perp}$
and parallel $k_{\parallel}$ to the line-of-sight.  The modulus of the
wave-number is then given by
$k=|\vec{k}|=\sqrt{k_\parallel^2+k_{\perp}^2}$.
The sensitivity at large $k$ depends crucially on the angular
resolution of the array (and therefore the maximum baseline), the
bandpass over which the measurement is made $\Delta \nu$, the
integration time $t_{\rm int}$, and the density of baselines which
measure a particular $k$-mode.  The noise may be divided into two
distinct components.  The thermal noise component $\Delta P_{21,N}$ is
proportional to the sky temperature, where $T_{\rm sky} \sim 250 \left(
\frac{1+z}{7}\right)^{2.6}$\,K at the frequencies of interest, and may
be written
\begin{equation}\label{thermal_noise}
\Delta P_{21,N}(\Vec{k}) = \left[\frac{T_{\rm sky}^2}{\Delta\nu t_{\rm int}} \frac{D^2 \Delta D}{n(k_{\perp})}\left(\frac{\lambda^2}{A_e}\right)^2\right] \frac{1}{\sqrt{N_c}},
\end{equation}
where $D$ is the co-moving distance to the centre of the survey volume
which has a co-moving depth $\Delta D$. Here $n(k_{\perp})$ is the
density of baselines which observe a wave-vector with transverse
component $k_{\perp}$. In the denominator $N_c$ denotes the number of
modes observed in a $k$-space volume $d^3k$. In terms of the
$k$-vector components, $N_c = 2 \pi k_{\perp} \Delta k_{\perp} \Delta
k_{\parallel} \mathcal{V}/(2 \pi)^3$ where $\mathcal{V} = D \Delta D
(\lambda^2/A_e)$ is the observed volume. Note that this calculation of
$N_c$ incorporates the fact that $k$-modes which do not fit in the
bandpass $\Delta \nu$ are not observable. The sample variance
component of the noise is given by $\Delta P_{21,SV}(\vec{k}) =
P_{21}(\vec{k})/\sqrt{N_c}$.  The sensitivity will improve with the
square root of the number of independent fields observed. The total
noise may therefore be expressed by
\begin{equation}\label{2dnoise}
\Delta P_{21}(k_{\perp},k_{\parallel},z) = \frac{\Delta
P_{21,N}(k_{\perp},k_{\parallel},z) + \Delta
P_{21,SV}(k_{\perp},k_{\parallel},z)}{\sqrt{N_{\rm fields} B_{\rm
tot}/\Delta \nu}}.
\end{equation}
For the purposes of constraining the parameters in
equation~\eqref{model1}, we average equation~\eqref{2dnoise} over the
viewing angle to obtain the spherically averaged noise $\Delta
P_{21}(k,z)$.  Note that this calculation assumes perfect foreground
removal over a bandpass with width $\Delta \nu$ [see
\citet{mcquinn2006} for a discussion of the relationship between
foreground removal and bandpass width].  In Section~\ref{fits} we
explore the effect of the choice of $\Delta \nu$ on the errors in the
measurement of $\alpha$.

We assume that the spherically averaged PS is measured in $n_k = 32$
logarithmically spaced intervals with end points in the range
0.05\,Mpc$^{-1} \leq k \leq 0.2$\,Mpc$^{-1}$.  Wave-numbers below this
range would not fit within the nominal bandwidth of $\sim8$\,MHz at
the redshifts of interest. Above $0.2$\,Mpc$^{-1}$ the BAO signal
becomes very weak and the matter PS may be non-linear.  The binning in
$k$-space, which is approximately equivalent to bin widths of $\Delta
k \sim k/20$, is motivated by the scale of the expected BAO
features. For clarity the grey error bars in Figures~1 and 2 represent
the uncertainty on the spherically averaged PS for larger spectral
bins with width $\Delta k = k/10$. We assume that three fields are
observed ($N_{\rm fields} = 3$) for 1000\,hrs each. Given the field of
view at these redshifts, this would correspond to surveying a few per
cent of the sky.

\section{Fits to 21\,cm power spectra}\label{fits}

In this section we describe the results of fits to the modelled 21\,cm
PS using the parametrized form in equation~\eqref{model1}.

\subsection{$\mathbf{\chi^2}$ minimization}

We adopt the $\chi^2$ function as an indicator for the ``goodness of fit'' of our model,
\begin{equation}
\chi^2(\Vec{p}) = \sum_{i=1}^{n_k}  \bigg[ \frac{P_{21}(k,z) - P_{\rm 21,fit}(\Vec{p},k,z)}{\sigma_{i}} \bigg]^2,
\end{equation}
where $\sigma_{i} = \Delta P_{21}(k,z)$ and $\Vec{p} \equiv \{ p_j \}
= (\alpha,b_0,b_1,b_2,b_3,c_0,c_1,c_2)$ is a vector containing the
model parameters. The index $i$ refers to the $n_k$ discrete values of
$k$ corresponding to each data bin (see Section~\ref{experiment}). We
restrict our fitting to the range $0.05\,$Mpc$^{-1}\le
k \le 0.2\,$Mpc$^{-1}$.

We have adapted the Numerical Recipes routines \citep{num_rec} which
implement the Levenberg-Marquardt technique to iteratively find the
parameter set which minimizes the $\chi^2$ function for a model which
is non-linear in its parameters $\vec{p}$. In Figures~1 and 2 we show
the resulting fits for the cases with a third order polynomial bias
(black lines), with and without a quadratic anomalous power term
respectively. The panels in the second column of Figure~1 (and the
first column in Figure~2) show the fitted spherically averaged 21\,cm
PS as well as the linear matter PS. The panels in the third column of
Figure~1 (and second column of Figure~2) show the fitted 21\,cm
bias. In Figure~2 the panels in the third column show the fitted $AP$
term (note that the input term is zero since non-linear effects are
not incorporated into the semi-numeric code we have used to generate
the ionisation maps).  The far right-hand panels in each of Figures~1
and 2 show the BAO component of the fitted PS, $B^2(k) D^2(z) P_l(k) +
AP(k) - b_{21}^2(k) D^2(z) P_{l,nb}(k)$. The noise in the fitted BAO
component is due to the noise in the model calculation of the bias
$b_{21}$.

\begin{figure*}
\begin{center}
	\begin{tabular}{cc}
    \epsfig{figure=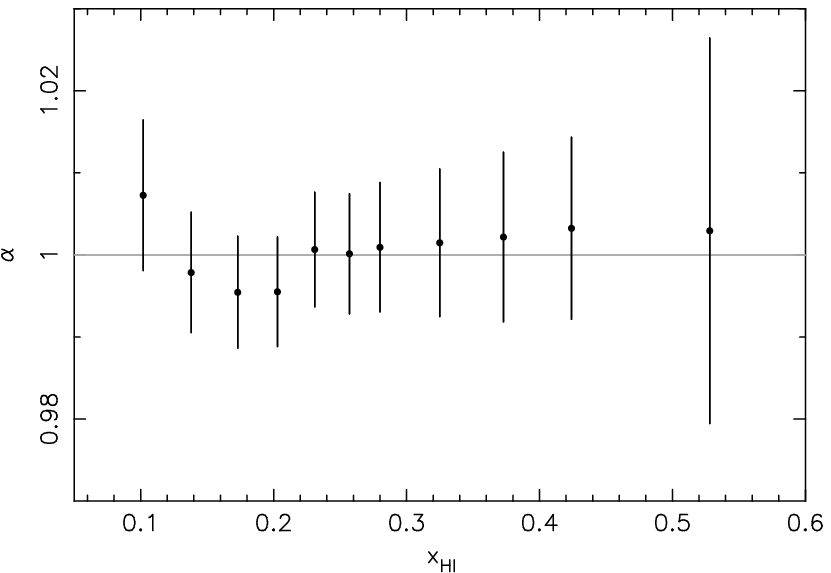,width=0.4\textwidth,angle=0,clip=}&
    \epsfig{figure=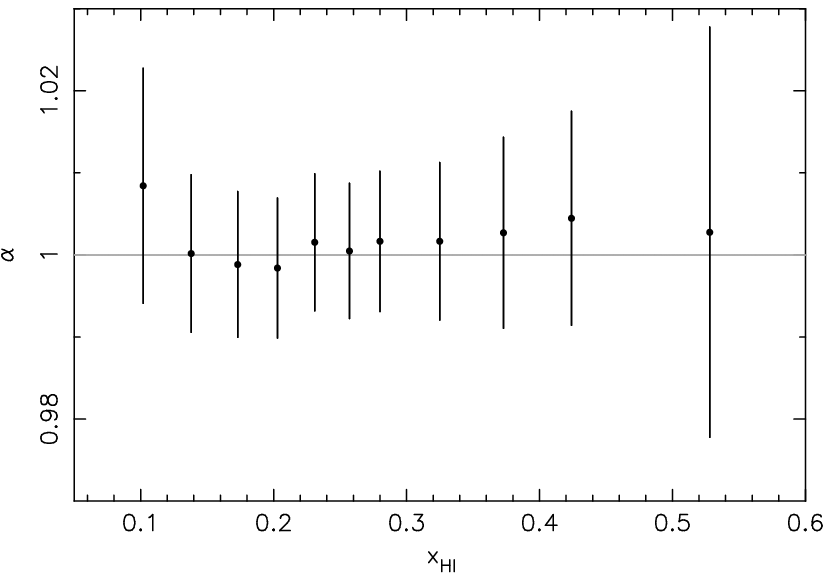,width=0.4\textwidth,angle=0,clip=}
  \end{tabular} \caption{\label{fig3}Fitted values and uncertainty for the $\alpha$ using a fit with third-order polynomial bias with (\emph{left panel}) and without (\emph{right panel}) a quadratic anomalous power term. The error bars represent the 1$\sigma$ ($68.3$~per~cent) confidence intervals calculated using from the Fisher matrix by marginalising over all of the other parameters in the fit.} 
\end{center}
\end{figure*}

The fit to the 21\,cm PS is very good in all cases.
When the anomalous power term is included there is a trade-off between
the power in the 21\,cm PS associated with the bias $B$ and that due
to $AP$, indicating some degeneracy. However as we show in the next
subsection, this degeneracy does not significantly bias the
constraints on the BAO scale. It may of course be possible to
alleviate the degeneracies between the bias and the $AP$ parameters by
putting some physically/numerically motivated priors on the values of
$c_0,c_1$ and $c_2$.

\subsection{Uncertainty in the measured BAO scale}\label{error}

\begin{figure*}
\begin{center}
	\begin{tabular}{cc}
    \epsfig{figure=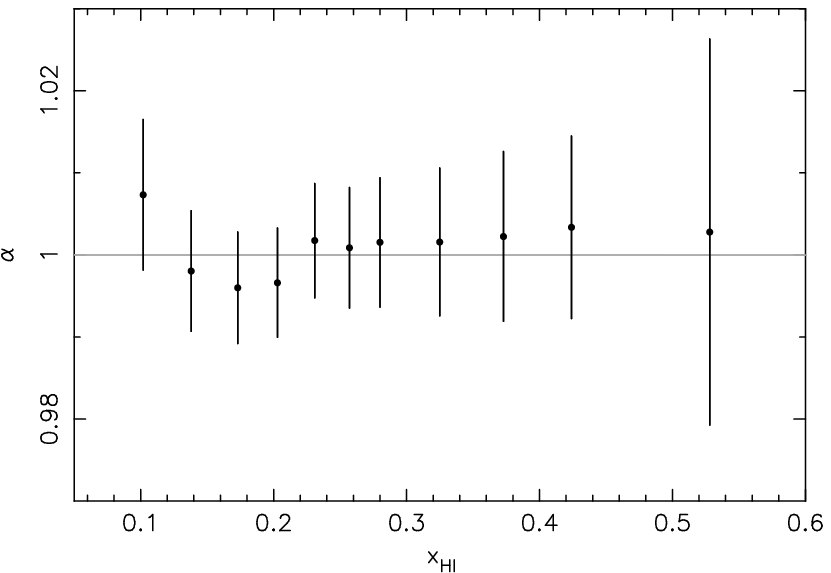,width=0.4\textwidth,angle=0,clip=}&
    \epsfig{figure=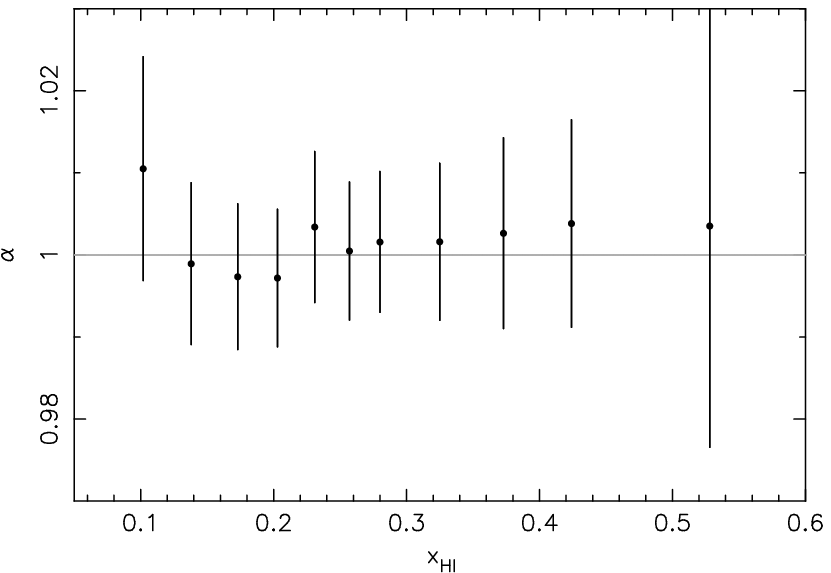,width=0.4\textwidth,angle=0,clip=}
  \end{tabular} \caption{\label{fig4}Fitted values and uncertainty for the $\alpha$ using a fit with fourth-order polynomial bias with (\emph{left panel}) and without (\emph{right panel}) a quadratic anomalous power term. The error bars represent the 1$\sigma$ ($68.3$~per~cent) confidence intervals calculated using from the Fisher matrix by marginalising over all of the other parameters in the fit.} 
\end{center}
\end{figure*}

After marginalising over the remaining fit parameters, we may
interpret the error in $\alpha$ for our fits as the fractional
uncertainty in the measurement of the BAO scale \citep{seo2005}.  The
Fisher matrix $\mathbf{F}$ is given by \citep{tegmark1997}
\begin{equation}
F_{ij} = \sum_{k=0}^{n_p} \sum_{l=0}^{n_p}\frac{\partial P_{\rm 21,fit}}{\partial p_i}\frac{\partial P_{\rm 21,fit}}{\partial p_j},
\end{equation}
where $n_p$ is the dimension of $\vec{p}$ and all derivatives are
evaluated using the best-fitting $\vec{p}$. Since we expect the noise
in 21\,cm PS to be Gaussian, we may invoke the Cramer-Rao theorem to
estimate the 1$\sigma$ ($68.3$~per~cent) confidence interval around
the best-fitting value of $\alpha$ from $\bf{F}$.  Assuming that all
other parameters are marginalised over this implies errors around the
best-fitting values for $\alpha$ of
\begin{equation}\label{error_eqn}
\Delta \alpha = (F_{11}^{-1})^{-1/2}.
\end{equation}

In Figure~\ref{fig3} we plot the the best-fitting values of $\alpha$
and the associated 1$\sigma$ error as calculated in
equation~(\ref{error_eqn}) for several redshifts in the range $6.4
\leq z \leq 8$ corresponding to average neutral fractions of $0.1
\lesssim x_{\rm HI} \lesssim 0.5$ for our reionisation model. The
\emph{left panel} of Figure~\ref{fig3} assumes a fitted 21\,cm PS with
no anomalous power term, whereas the fitted PS used to generate the
plot in the \emph{right panel} includes the $AP$ term. In each panel
the grey horizontal line indicates the input value $\alpha = 1$. Note
that the recovered values of $\alpha$ deviate from the input values at
$x_{\rm HI} \sim 0.1 - 0.2$ ($z \sim 6.4 - 6.7$).  However for the
observing strategy and $k$-space binning chosen, this deviation is
within the estimated $1\sigma$ statistical error (although only just
at $z\sim 6.4$ where the neutral fraction is $\sim 10$~per~cent).  As
reionisation nears completion ($x_{\rm HI} \rightarrow 0.1$), the
21\,cm bias becomes most strongly scale dependent in the range
0.05\,Mpc$^{-1} \lesssim k \lesssim 0.2$\,Mpc$^{-1}$. The increasing
systematic error toward $z \sim 6.4$ is therefore a result of the
(relatively) low noise level in the 21\,cm PS at these redshifts
combined with the increasingly strong evolution of the bias function
within the fitted range of $k$-values.

We find that the statistical errors are slightly larger when we
include the extra parameters required to describe the anomalous power
(since the matrix $\mathbf{F}$ is not diagonal). We note that our
estimate of the attainable errors on the phase/dilation parameter
$\alpha$ are a minimum in the sense that $(i)$ we have assumed perfect
foreground removal across each bandpass with width 8\,MHz for a total
bandwidth of 32\,MHz, and $(ii)$ we have invoked the Cramer-Rao
theorem regarding the minimum 1$\sigma$ error on a parameter from the
Fisher matrix of the best-fitting model.

The bandpass over which foreground removal can be performed may affect
the errors on the model parameters, even for a fixed total bandpass
$B_{\rm tot}$, because the removal of foreground on larger scales
allows for more of the strong large-scale (small $k$) BAO peaks to be
included in the fit. However we find that at the high redshifts that
we are considering, increasing $\Delta \nu$ above 8\,MHz has little
affect on the errors in the recovered values of $\alpha$. On the other
hand, we find that decreasing $\Delta \nu$ increases the systematic
uncertainty in the recovered values of $\alpha$ toward the end of
reionisation. For model fits without the $AP$ term, the systematic
error exceeds the statistical error when $x_{\rm HI} \lesssim 0.2$
(corresponding to redshifts $z \lesssim 6.7$), indicating that our
maximum likelihood indicator is biased. This highlights the
potentially significant advantages of including the larger scales
accessible with wider bandpasses $B$.

The systematic component of the errors will in general be sensitive to
the functional form chosen to describe the 21\,cm PS. We find that the
effect of increasing the order of the polynomial used to fit the bias
is somewhat degenerate with the chosen value of $n_k$.  For example,
with coarser k-space binning of $n_k = 16$ the errors in $\alpha$
decrease by $\sim 0.5$~per~cent when the bias is modelled by a fourth
order polynomial rather than a third order polynomial. This effect is
not observed for $n_k = 32$ (see Figures~\ref{fig3} and
\ref{fig4}). This implies that sampling of the BAO features must be
sufficiently fine. However, the affect of $n_k$ on the systematic
errors in $\alpha$ is only significant for $x_{\rm HI} < 0.2$ when the
bias is evolving strongly over the range of fitted k-values.

\section{Summary and conclusions}\label{summary}

There has been recent interest in the utility of redshifted 21\,cm
fluctuations from the IGM as a probe of the mass PS at a range of
redshifts \citep[e.g.][]{mcquinn2006, bowman2006,
mao_tegmark2008,wlg2008,chang2008,mao2008}. During the reionisation
era ($z\ga6$), the appearance of HII regions generates scale
dependence of the bias relating the 21\,cm PS to the linear PS. In
this paper we have set out to test the extent to which the scale
dependence of the 21\,cm bias will interfere with the ability to
extract the BAO scale from the PS of 21\,cm fluctuations.

We have utilised the results of semi-numerical models for the
evolution of the ionised hydrogen distribution on large scales in
order to construct an estimate for the 21\,cm PS at high redshift. For
a model in which the IGM is fully reionised by $z=6$, and assuming the
design parameters for a 10-fold expansion the the MWA, we find that
the scale dependent 21\,cm bias limits the precision, but not the
accuracy, of measurements of the BAO scale in the window $6.5 \lesssim
z < 8.0$. We parametrize the bias and anomalous power terms in the
21\,cm PS as polynomials, and for the telescope design assumed, find
errors on the spherically averaged BAO scale of $\lesssim
1.5$~per~cent in the window $6.5 \le z \le 7.5$. At redshifts closer
to the end of reionisation, the strong scale dependence of the 21\,cm
bias introduces systematic error in the BAO scale, which exceeds the
level of the statistical uncertainty for narrower bandpasses ($B <
8~{\rm MHz}$). We find that the sensitivity of a low-frequency array
to the spherically averaged BAO scale decreases with increasing
redshift, becoming $> 2$~per~cent by $z \sim 8$.

The details of the quantitative results presented in this paper are
sensitive to the reionisation model assumed. More generally, the
constraints available at a particular redshift will be sensitive to
the details of the reionisation history. The best constraints are
likely to be derived from observations of the Universe when the IGM is
$\sim50$~per~cent ionised - at which time the fluctuations are
maximised \citep{lidz2008} - but the bias is relatively weakly
dependent on scale at the scale of the BAO signal (and hence does not
introduce a large systematic error). As a result, if reionisation
completes earlier than $z \sim 6$, the most precise constraints will
be obtained at higher redshift than reported here. However, as the
foregrounds become brighter when observing at higher redshift, an
early reionisation scenario would decrease the maximal precision with
which the BAO scale could be measured.

A detailed analysis of the corresponding constraints on the dark
energy equation of state is non-trivial and depends on the functional
form assumed \citep{wlg2008}. We do not explore the translation of BAO
scale to cosmological constraints in this paper. Such an analysis
should include independent complementary constraints (e.g. from the
CMB), and take account of uncertainties in, for example, the horizon
scale \citep[e.g.][]{mcquinn2006,mao_tegmark2008}. Simple estimates of
the cosmological constraints from a high redshift BAO measurement were
presented in \citet{wlg2008}.

In a recent paper \citet{mao_tegmark2008} propose a parametrization of
the 3d 21\,cm PS including peculiar velocity effects which is
motivated by the generic form expected for the ionisation PS and
ionisation-matter PS. Our analysis of the accuracy with which the BAO
scale may be recovered in the presence of a scale dependent 21\,cm
bias is complementary to their detailed discussion of the cosmological
constraints available via the redshifted 21\,cm
PS. \citet{mao_tegmark2008} demonstrate that their parametrization can
recover the ionisation PS constructed from simulations of reionisation
and therefore can be used to constrain cosmological parameters from
the 21\,cm PS. Whilst they mention that their parametrization
struggles to describe the high-$k$ ionisation PS by a redshift of
$z=7$ when the ionisation PS become significantly scale dependent,
this is not found to be a problem up to the $k_{\rm max} =
0.2$\,Mpc$^{-1}$ that we consider in this paper.  It would be
interesting to see how well such a parametrization would perform at
the end of reionisation in the wavelength range that we have used for
our fits.

Our calculations of the constraints on the BAO scale from 21\,cm
intensity fluctuations are promising when compared to the expected
performance of future galaxy surveys. \citet{glazebrook2005} show that
a dedicated next-generation spectroscopic galaxy survey could measure
the BAO scale to an accuracy of $\sim 1$~per~cent for the transverse
component at a redshift $z \sim 3.5$, with a slightly reduced
sensitivity to the line-of-sight BAO scale. Limitations in the
precision achievable arise due to the large number of galaxies needed
to reduce the cosmic variance on the scales at which the BAO signal is
expected ($\gtrsim 30$\,Mpc), combined with the decrease in the number
of sufficiently luminous galaxies toward higher redshift. The accuracy
with which the BAO scale may be measured via the 21\,cm intensity PS
is immune to both of these constraints because there is no need to
resolve individual objects. Redshifted 21cm measurements of the PS will
instead be limited by cosmic variance owing to the finite size of the
field of view, and sensitivity to individual modes owing to array
configuration and finite collecting area. However we find that the PS of
redshifted 21\,cm fluctuations as measured by the MWA5000 could
achieve constraints on the BAO scale at $z>6$ that are comparable to
the accuracy which will be available to future galaxy redshift surveys
at $z\la3.5$.

In summary, one potential obstacle to measuring the BAO scale in
21\,cm fluctuations during the reionisation era is the strongly scale
dependent bias between the 21\,cm and mass PS that is
produced by the appearance of large ionised regions. In this paper we
have shown that the appearance of ionised regions during the
reionisation epoch will not inhibit the measurement of the BAO scale
during most of the reionisation era.

\section*{Acknowledgments} We thank the referee for useful
comments. KJR acknowledges the support and hospitality of the
Department of Physics, University of Melbourne, and a visiting
postgraduate Marie Curie Fellowship hosted by MPA. The research was
supported by the Australian Research Council (JSBW). PMG acknowledges
the support of an Australian Postgraduate Award.

\bibliographystyle{mn2e} \bibliography{citations}

\end{document}